\documentclass[12pt]{article}%

\usepackage{graphicx}
\usepackage{amsthm}
\usepackage{amsmath}
\usepackage{amsfonts}
\usepackage{amssymb}
\usepackage{color}
\usepackage[normalem]{ulem}
\usepackage{mathrsfs}

\newcommand{\U}{\mathscr{U}}
\newcommand{\C}{\mathcal{C}}
\newcommand{\N}{\mathbb{N}}
\newcommand{\E}{\mathbb{E}}

\newcommand{\Ec}[2]{\E \Big[#1\Big| #2\Big]}
\renewcommand{\P}{\mathbb{P}}

\newcommand{\eh}{\hat{\eta}}
\newcommand{\codi}{\Rightarrow}
\newcommand{\sL}{\sqrt{2\lambda}}

\newcommand{\LT}[1]{{\mathbb{CT}^{(#1)}}}
\newcommand{\LTT}{{\mathbb{CT}}}
\newcommand{\TT}[1]{{\mathbb{T}^{(#1)}}}
\newcommand{\TTT}{{\mathbb{T}}}
\newcommand{\TR}{{\mathcal{T}}}
\newcommand{\G}{{\mathcal{G}}}
\newcommand{\GG}{{\mathcal{G}^\infty}}
\newcommand{\gen}{\operatorname{gen}}
\newcommand{\dist}{\operatorname{dist}}

\numberwithin{equation}{section}

\newtheorem{theorem}{Theorem}[section]

\newtheorem{corollary}{Corollary}[section]

\theoremstyle{definition}
\newtheorem{definition}{Definition}

\theoremstyle{remark}
\newtheorem{remark}{Remark}[section]

\begin{document}

\author{
V.~Sisko\footnote{Department of Mathematics, Universidade Federal Fluminense, Brazil}, A.~Yambartsev\footnote{Department of Statistics, University of S\~ao Paulo, Brazil} and S.~Zohren\footnote{Rudolf Peierls Centre for Theoretical Physics, Oxford University, UK; work done while author was at the Mathematical Institute, Leiden University, The Netherlands and the Department of Statistics, University of S\~ao Paulo, Brazil}
}
\title{A note on weak convergence results for uniform infinite causal triangulations}
\date{December 1, 2011}
\maketitle

\begin{abstract}
We discuss uniform infinite causal triangulations and  equivalence to the size biased branching process measure - the critical Galton-Watson branching process distribution conditioned on non-extinction. Using known results from the theory of branching processes, this relation is used to prove weak convergence of the joint length-area process of a uniform infinite causal triangulations to a limiting diffusion. The diffusion equation enables us to determine the physical Hamiltonian and Green's function from the Feynman-Kac procedure, providing us with a mathematical rigorous proof of certain scaling limits of causal dynamical triangulations. \\ \\
\textbf{2000 MSC.} 60F05, 60J60, 60J80.\\
\textbf{Keywords.} Causal triangulation, scaling limits, weak convergence, diffusion process, branching process. \\ \\
\emph{Submitted to MPRF in 12/2010, revised version 12/2011}
\end{abstract}
\newpage
\section{Introduction}

Models of planar random geometry provide a rich field with an interplay between mathematical physics and probability.

On the physics side so-called dynamical triangulations (DT) have been
introduced as models for two-dimensional Euclidean quantum gravity and string
theory (see e.g. \cite{Ambjorn:1997di} for an overview). The basic idea is to
define the gravitational path integral as a sum over triangulated surfaces. Any
physical observable is then defined on the ensemble of all such triangulations.
At the end, continuum physics is obtained by performing a scaling limit in
which one takes the size of the triangulations to infinity keeping the physical
area constant.

On the probabilistic side Angel and Schramm \cite{Angel:2002ta} first
introduced the uniform measure on infinite planar triangulations proving the
existence of the above scaling limit as a weak limit. This construction was
essential to prove several properties of such uniform infinite triangulations.
In particular Angel \cite{Angel2003} proved that the volume of a ball $B(R)$ of
radius $R$ is of order $R^{4}$ and that the length of the boundary is of order
$R^2$. This proved rigorously that the fractal dimension of such triangulations
is $d_H=4$, a result long known to physicists (see e.g. \cite{Ambjorn:1997di}).
Later Krikun \cite{Krikun2005} obtained the exact limit theorem for the scaled
boundary length of $B(R).$

\begin{figure}[ptb]
\begin{center}
\includegraphics{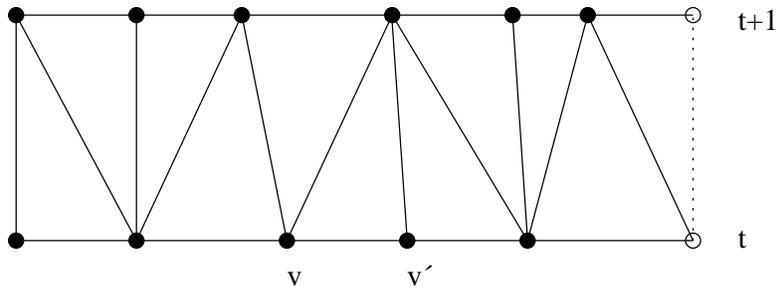}
\end{center}
\caption{A time slice from time $t$ to time
$t+1$. The left and right side of the strip should be identified to form a band with topology
$S^1\times[t,t+1]$.}%
\label{f0}
\end{figure}

While two-dimensional Euclidean quantum gravity defined through DT definitely
has a rich mathematical structure as pointed out above, as a model of quantum
gravity it failed to be numerically extended to higher dimensions. This lead to
the development of a different approach of so-called Causal Dynamical
Triangulations (CDT) by Ambj{\o}rn and Loll \cite{Ambjorn:1998xu}. In contrast
to the \emph{Euclidean} model, CDT provides a nonperturbative definition of the
\emph{Lorentzian} gravitational path integral. These causal triangulations
differ from their Euclidean analogs in the fact that they have a time-sliced
structure of fixed spatial topology. Consider for example triangulations of an
overall topology of a cylinder. Then the triangulation consists of slices
$S^1\times[t,t+1]$ from time $t$ to time $t+1$ as illustrated in
Figure~\ref{f0}. Here, edges connecting vertices in slices of equal time are
called space-like edges, while edges connecting subsequent slices are called
time-like edges.

Note that this class of triangulations forms a causal structure needed to model
Lorentzian geometries. In particular, we can think that a vertex $v'$ lies in
the future of a vertex $v$ iff there is a path of time-like edges leading from
$v$ to $v'$. For example, the vertices $v$ and $v'$ as illustrated in
Figure~\ref{f0} are not causally related.

The physical properties of the ensemble of causal triangulations behaves much
more regular than its Euclidean counterpart. For example, it has a fractal
dimension of $d_H=2$ instead of $d_H=4$ for DT. Also when coupled to simple
matter models, such as the Ising model, it behaves much more like a regular
lattice such as $\mathbb{Z}^2$ \cite{Ambjorn:1999gi}.

While the approach of CDT has recently lead to a number of interesting physical
results, especially with respect to higher-dimensional numerical
implementations (see \cite{Ambjorn:2010ce} for a review), the probabilistic
aspects of this model have hardly been studied. In fact, recently
Durhuus, Jonsson and Wheater defined the uniform measure on infinite
causal triangulations \cite{Durhuus:2009sm} (see also \cite{DurhuusZakopane}
for earlier ideas), proving almost surely (a.s.) recurrence, that a.s.\ the
fractal dimension is $d_H=2$ and that the spectral dimension is a.s.\ bounded
from above by $d_s\leq 2$. A similar definition of the uniform measure has
previously also been used by one of the authors of this article and M. Krikun
to prove the existence of a phase transition for the Ising model coupled to CDT
\cite{anatoli}.

In this article we discuss the existence of the uniform
measure on infinite causal triangulations in an alternative presentation to
\cite{Durhuus:2009sm}. In the line of \cite{Angel:2002ta} and \cite{Krikun2005}
we give weak convergence limits for the distribution of the area and length of
the boundary of a ball of radius $t$, confirming that they scale as $t^2$ and
$t$. These results follow from a bijection between causal triangulations and
certain Galton-Watson branching processes through the size biased branching
process measure \cite{Lamperti1968} - the critical branching process
distribution conditioned on survival at infinity. Exploiting the relation to
conditioned critical Galton-Watson processes, one can go further and obtain
weak convergence of the joint length and area process. The process is diffusive
and the corresponding Kolmogorov equation enables us to derive the physical
Hamiltonian, providing us with a mathematical rigorous formulation of scaling
limits of CDT. \\

In the next section we give basic definitions and introduce infinite causal
triangulations and show existence of the uniform measure on infinite causal
triangulations in an alternative presentation to \cite{Durhuus:2009sm}.
In Section \ref{branching} we then present the relation to
critical Galton-Watson processes conditioned to never die out. In Section
\ref{distribution} we exploit this relation to obtain weak convergence of the
length process (Theorem \ref{thmlength}) and the joint length-area process
(Theorem \ref{thmarea}). Theorem  \ref{thmlength} is proven in Appendix
\ref{markov}. These results provide a mathematical rigorous formulation  of
certain scaling limits of CDT which we discuss in Section \ref{conclusion}.

\section{Uniform infinite causal triangulations}\label{UnifSec}

We consider rooted causal triangulations of
$\C_h=S^1\times [0,h]$, $h=1,2,\dotsc$,
and  of $\C=S^1\times[0,\infty)$.

\begin{definition}
Consider a (finite) connected graph $G$. Its embedding $i:G\to \overline{\C_h}$ is called
a \emph{causal triangulation} $T$ of $\C_h$
if the following conditions hold:
\begin{itemize}
\item the (open) disks attached to $\C_h$ in order to get $\overline{\C_h}$ are faces of $T$, and all faces, with possible exception of these two disks, are triangles;
\item each face of $T$ that belongs to $\C_h$ belongs to some strip $S^1\times[j,j+1],
j=0,1,\dots,h-1$ and has all vertices and exactly one side on the
boundary $(S^1\times\{j\})\cup (S^1\times\{j+1\})$ of the strip
$S^1\times[j,j+1]$.
\end{itemize}
\end{definition}

\begin{remark}
 Some care has to be put into the definition of what is meant by a triangle due to self-loops and multiple edges, and in particular, a simple definition like
``the face is a triangle if its boundary meets precisely three edges of the
graph'' is not enough. Let the size of a face be the number of edges incident
to it, with the convention that an edge incident to the same face on both sides
counts for two. We then call a face with size 3 (or $3$-sided face) a \emph{triangle}.
\end{remark}

\begin{definition}
\label{def:rct}
A causal triangulation $T$ of $\C_h$ is called \emph{rooted} if it has a root.
The \emph{root} $(x,e)$ of $T$ consists of
vertex $x$ and directed edge $e$ that runs from $x$, they
are called \emph{root vertex} and \emph{root edge} correspondingly.
The root vertex and the root edge belong to $S^1\times \{0\}$.
The orientation induced by the ordered pair that consists of the root edge and
the vector that runs from the root vertex in positive time direction
coincides with the fixed orientation of $\overline{\C_h}$.
\end{definition}

\begin{definition}
\label{def:rcte}
Two rooted causal triangulations of $\C^h$, say $T$  and
$T'$, are \emph{equivalent} if the following conditions hold:
\begin{enumerate}
\item $T$ and $T'$ are embeddings of the same graph $G$, that
is, $T$ is embedding $i:G \to \overline{\C^h}$ and
$T'$ is embedding $j:G \to \overline{\C^h}$;
\item there exists a
self-homeomorphism $\tilde h:\overline{\C^h}\to \overline{\C^h}$ such that $\tilde h i=j$.
Here we suppose that $\tilde h$ not only takes $i(G)$ to $j(G)$ but also
transforms each slice
$S^1\times\{j\}$, $j=0,\dotsc,h$ to itself and
sends the root of $T$ to the root of $T'$.
\end{enumerate}
\end{definition}

For convenience, we usually abbreviate ``equivalence class of
rooted causal triangulations'' to ``causal
triangulation'' or CT.

Cutting off the stripe $S^1\times (h,h+1]$ from $\C^{h+1}$ we obtain a natural
map from the set of causal triangulations of $\C^{h+1}$ to the set of causal
triangulations of $\C^h$ that we denote by~$r_h$.
\begin{definition}
We say that $T$ is a causal
triangulation of $\C$, if $T=(T_1,T_2,\dotsc)$,
where $T_h$ is a causal triangulation of $\C_h$, $h=1,2,\dotsc$,
and
the sequence is subject to consistency condition  $T_h=r_h(T_{h+1} )$, $h=1,2,\dotsc$.
\end{definition}


By $\LTT_\infty$ denote the set of all causal triangulations of $\C$ and by
$\LTT_h$ denote the set of all causal triangulations of $\C^h$.
Let
\[
\LT{h}=\bigcup_{i=1}^h \LTT_i, h=1,2,\dotsc
\quad
\text{and}
\quad
\LT{\infty}=\LTT_\infty\cup \bigcup_{i=1}^\infty \LTT_i.
\]
The restriction map $r_h\colon \LTT_{h+1} \to \LTT_h$ can be naturally generalized to become
the restriction map $r_h\colon \LT{\infty} \to \LT{h}$.
We see that $T\in \LT{\infty}$ is identified by the sequence
$(T_1,T_2,\dotsc)$,
where $T_h\in \LT{h}$ are
subject only to consistency condition  $T_h=r_h(T_{h+1} )$, $h=1,2,\dotsc$.

We use the standard formalism for plane trees (see \cite{N1986} or \cite{Aldous1998}).
Let
$${\U}=\bigcup_{n=0}^\infty \N^n  $$
where $\N=\{1,2,\ldots\}$ and by convention $\N^0=\{\varnothing\}$.
The \emph{height} of $u=(u_1,\ldots,u_n)\in\N^n$ is $|u|=n$. If
$u=(u_1,\ldots u_m)$ and
$v=(v_1,\ldots, v_n)$ belong to $\U$, then $uv=(u_1,\ldots u_m,v_1,\ldots ,v_n)$
denotes the concatenation of $u$ and $v$. In particular $u\varnothing=\varnothing u=u$.
If $v$ is of the form $v=uj$ for $u\in{\U}$ and $j\in\N$,
we say that $u$ is the {\it predecessor} of $v$, or that $v$ is a {\it successor} of $u$.
More generally, if $v$ is of the form $v=uw$ for $u,w\in{\U}$, we say that
$u$ is an {\it ancestor} of $v$, or that $v$ is a {\it descendant} of $u$.

\begin{definition}
\label{df:ftree}
A (finite or infinite) \emph{family tree} $\tau$ is a subset of
$\U$ such that
\begin{description}
\item{(i)} $\varnothing\in \tau$;

\item{(ii)} if $u\in \tau$ and $u\ne\varnothing$, the predecessor of
$u$ belongs to $\tau$;

\item{(iii)} for every $u\in\tau$, there exists an integer $k_u(\tau)\geq 0$
such that $uj\in\tau$ if and only if $1\leq j\leq k_u(\tau)$.
\end{description}
\end{definition}
A \emph{height of a finite tree} is the maximum height of all vertices in the tree.
Let $\TT{\infty}$ be the set of all family trees and $\TT{h}$
the set of all finite family trees of height at most $h$.
There is a natural restriction map $r_h:\TT{\infty}\to \TT{h}$ such that
 if $\tau$ is a family tree, then
$r_h \tau$ is the tree formed by all vertices of $\tau$ of height at most $h$.
Let $\TTT_\infty$ be the set of all infinite family trees
and $\TTT_h$ the set of family trees of height $h$.

A family tree $\tau\in \TT{\infty}$ is identified by the
sequence $(r_h\tau,h\ge 1)$. Note that the $r_h\tau \in \TT{h}$ are subject
only to the consistency condition that $r_h\tau=r_h(r_{h+1}\tau)$.

\begin{figure}[t]
\begin{center}
\includegraphics[width=13cm]{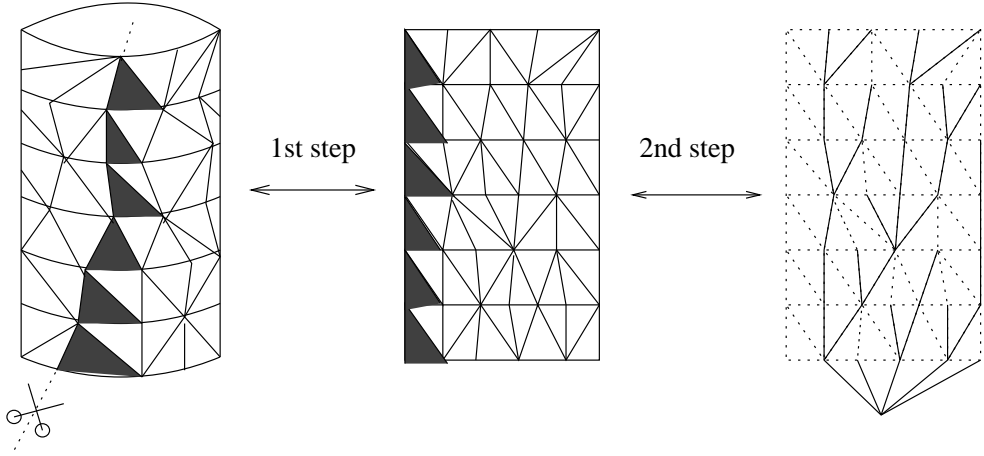}
\end{center}
\caption{Tree parametrization. The following two steps outline how to go from a causal triangulation to a tree: {\it Step 1.} The first step is cutting. We construct the sequence of root vertices (or triangles) on each slice
$S^1\times\{i\}$ by the following rule. Let $v_1$ be the vertex in the
slice $S^1\times\{1\}$ which belongs to the rooted triangle
containing the root edge $[v_0,v_0'].$ One chooses the right most
neighbour $v_1'$ of $v_1$ on the slice $S^1\times\{1\}$ as the new rooted edge. Following this procedure we can cut open the triangulation along the left side of the root triangles. {\it Step 2.} We add one vertex in the slice below the initial boundary and connect all vertices on the inital boundary to this vertex. We then remove all spatial (horizontal edges) and each leftmost outgoing time-like edge of every vertex. The lowest vertex is then connected to the root.  The resulting graph is a tree. The inverse relation should now be clear from the construction. }%
\label{f2}%
\end{figure}

\begin{theorem}
There is a bijection  $\phi\colon\LT{\infty}\to \TT{\infty}$ such that
\begin{itemize}
\item $\phi\circ r_h=r_h\circ \phi$, that is,  $\phi$ respects $r_h$, $h=1,2,\dotsc$;
\item for $t=1,2,\dotsc,\infty$, restrictions of $\phi$ to $\LTT_t$ denoted by
$\phi_t\colon \LTT_t\to \TTT_t$ are also bijections
that respect $r_h$, $h=1,2,\dotsc$.
\end{itemize}
\end{theorem}
This theorem dates back to \cite{DiFrancesco:2000nn} and a detailed proof can be found in \cite{Durhuus:2009sm} (see also \cite{MYZ2001}). We refer to Figure \ref{f2} for an illustration of the proof. 

The set $\TT{\infty}$ is now identified as a subset of an infinite
product of countable sets
\[
\TT{\infty} \subset \TT{0}\times\TT{1}\times\TT{2}\times\dots
\]
We give $\TT{\infty}$ the topology derived by this identification
from the product of discrete topologies on $\TT{h}$. Therefore,
a sequence of family trees $\tau_n$ has a limit
$$ \lim \tau_n= \tau \in \TT{\infty}$$
iff for every $h$ there exist a $\tau^{(h)}\in \TT{h}$ and $n(h)$ such
that $r_h \tau_n=\tau^{(h)}$ for all $n\ge n(h)$; the limit is then the
unique $\tau\in \TT{\infty}$ with $r_h \tau=\tau^{(h)}$. In particular, for
each $\tau \in \TT{\infty}$ the sequence $r_h\tau$ has limit $\tau$ as $n\to
\infty$. The topology is metrizable, e.g., set $d(\tau,\tau')=k^{-1}$, where
\[
k=\sup\{h:r_h\tau=r_h\tau'\}.
\]
It is easy to see that the metric space is
complete and separable.

The topology gives us the Borel $\sigma$-algebra to define probability
measures on it. Besides, we can define the weak convergence of measures.
As usual, a measure $\mu$ is the weak limit of the sequence of
measures $\mu_n$ if
$$ \int f d\mu_n \to \int f d\mu, \text{ as } n \to \infty$$
for every bounded continuous real-valued function $f$ given on
$\TT{\infty}$.

Let $\TTT=\cup_{h=0}^{\infty}\TT{h}$.
Consider a system of nonnegative numbers $$\pi=\{p(\tau), \tau \in \TTT\}$$ such
that the following conditions hold:
\begin{enumerate}
\item for $h=0,1,2,\dots$ we have
\[
\sum p(\tau_{h+1})=p(\tau_h) \text{ for any } \tau_h \in \TT{h},
\]
where the sum is over $\tau_{h+1}\in \TT{h+1}$ such that
$r_h\tau_{h+1}=\tau_h$;
\item $p(\tau_0)=1$ for $\tau_0\in \TT{0}$.
\end{enumerate}
It is easy to see that if $\mu$ is a probability measure on
$\TT{\infty}$, then
\[
 \pi=\left\{\mu\bigl(\{\tau\in \TT{\infty}: r_h\tau=\tau_h\}\bigr): \tau_h \in \TT{h},
h=0,1,2,\dots\right\}
\]
is a system of numbers that satisfies the above two conditions.

The following fact can easily be checked (it is proved in the same
way as Kolmogorov extension theorem). It
helps to define a measure on $\TT{\infty}$.
The
fact is that for every
system of nonnegative numbers $\pi$ satisfying the two conditions
above there is a
probability measure $\mu$ on $\TT{\infty}$ such that
\[
 p(\tau_h)=\mu\bigl(\{\tau\in \TT{\infty}: r_h\tau=\tau_h\}\bigr) \text{ for all } \tau_h
\in \TT{h}, h=0,1,2,\dots.
\]
In other words, a random family tree is a random element
of $\TT{\infty}$, formally specified by its sequence of
restrictions, say $\TR=(r_h\TR, h=0,1,\dots)$, where
each $r_h\TR$ is a random variable with values in the
countable set $\TT{h}$, and
$r_h\TR=r_h(r_{h+1}\TR)$ for all $h$. The distribution
of $\TR$ is determined by the sequence of distributions of
$r_h\TR$ for $h\ge0$. Such a distribution is determined by
a specification of the conditional distributions of $r_{h+1}\TR$ given
$r_h\TR$ for $h\ge0$. To give a more exact specification of the distribution
of $\TR$, we need some definitions.

For every $v\in \U$, let $c_v \tau$ be the number of successors of $v$ (if
$v\notin \tau$, then $c_v \tau=0$).
For every $\tau \in \TT{\infty}$ and $g\ge0$, let the \emph{$g$th generation of
individuals in $\tau$}, denoted by $\gen(g,\tau)$, be the set of $u\in \tau$
such that the height of $u$ is $g$,
also let $Z_g\tau$ be the number of elements of the set $\gen(g,\tau)$ (to simplify notation let $Z=Z_0$).
Note that $\TT{0}$ contains only one family tree that consists of only one element $\varnothing$, and for any $\tau\in \TT{\infty}$, we have $r_0\tau=\{\varnothing\}$.
A family tree $\tau$ is conveniently specified as the unique $\tau\in
\TT{\infty}$ such that $r_h\tau=\tau^{(h)}$ for all  $h$ for some sequence of
trees $\tau^{(h)} \in \TT{h}$ determined recursively as follows. Given that
$\tau^{(h)} \in \TT{h}$ has been defined ($\tau^{(0)}$ is the unique tree from
$\TT{h}$), the set of vertices
$\gen(h,\tau)=\gen(h,\tau^{(h)})=r_h\tau\setminus r_{h-1}\tau$ is determined,
hence so is the size $Z_h \tau=Z_h \tau^{(h)}$ of this set; for each possible
choice of $Z_h \tau$ non-negative integers $\bigl(a_v, v \in
\gen(h,\tau)\bigr)$, there is a unique $\tau^{(h+1)} \in \TT{h+1}$ such that
$r_h\tau^{(h+1)}=\tau^{(h)}$ and $c_v \tau^{(h+1)}=a_v$ for all $v \in \gen(h,\tau)$.
So a unique $\tau\in\TT{\infty}$ is determined by specifying for each $h\ge 0$ the way in which these $Z_h \tau$ non-negative integers are chosen given that $r_h\tau=\tau^{(h)}$ for some $\tau^{(h)} \in \TT{h}$.

Thus a more exact specification of  the distribution of $\TR$
is
a specification of the joint conditional distribution given $r_h\tau$ of the
numbers of children $c_v\tau$ as $v$ ranges over $\gen(h,\tau)$ for $h\ge0$.

Since the topology on $\TT{\infty}$ is a product of discrete topologies on $\TT{h}, h\ge0$,
the weak convergence of measures on $\TT{\infty}$ can be easily reformulated in the following way.
For random family trees $\TR_n, n=1,2,\dotsc$ and $\TR$, we say that $\TR_n$
converges in distribution to $\TR$, and write $\dist(\TR_n)\to \dist(\TR)$ if
\[
\P(r_h \TR_n=\tau)\to
\P(r_h \TR=\tau) \quad \forall h \ge 0, \tau \in \TTT.
\]

\section{Uniform infinite causal triangulations and critical branching processes}\label{branching}
Let $p(\cdot)=(p(0),p(1),\dotsc)$ be a probability distribution on the non-negative integers with $p(1)<1$.
Call a random family tree $\G$ a \emph{Galton-Watson (GW) tree with offspring distribution} $p(\cdot)$ if the number of children $Z\G$ of the root has distribution $p(\cdot)$:
\[
\P(Z\G=n)=p(n) \quad \forall n \ge 0
\]
and for each $h=1,2,\dotsc$, conditionally given $r_h \G=t^{(h)}$, the numbers of children $c_v\G$, $v \in \gen(h,t^{(h)})$, are i.i.d.\ according to $p(\cdot)$.

Introduce the generating function $f(s)=\sum_{n\geq0} p(n) s^n$. We consider only the \emph{critical} GW process which has $f'(1)=1$. Suppose further that $\nu=f''(1)/2<\infty$.
A random family tree $\GG$, which we call $\G$ \emph{conditioned on
non-extinction} is derived from the family tree $\G$ in the way described in
Theorem~\ref{t:big} below. The probabilistic description of $\GG$ involves the
\emph{size-biased} distribution $p^*(\cdot)$ associated with probability
distribution $p(\cdot)$:
\[
p^*(n)=\mu^{-1}np(n) \quad \forall n\ge 0.
\]
Putting together Proposition 2 and Proposition 5 from \cite{Aldous1998} (they correspond to reformulation in this family tree language of theorems from \cite{Kesten1986} and \cite{Kennedy1975}), we have the following
\begin{theorem}
\label{t:big}
The following statements are valid.
\begin{enumerate}
\item
\begin{equation}\label{yamb:n1}
\dist(\G|\# \G=n) \to \dist(\GG) \quad \text{as $n\to \infty$},
\end{equation}
where  $\dist(\GG)$ is the distribution of a random family tree $\G^\infty$ specified by
\begin{equation}
\P(r_h\GG=\tau)=Z_h \tau \P(r_h\G=\tau) \quad \forall \tau \in \TT{h}, h\ge 0.
\end{equation}
\item
Almost surely $\G^\infty$ contains a unique infinite path $(V_0,V_1,V_2,\dotsc)$ such that $V_0=\varnothing$ and
$V_{h+1}$ is a successor of $V_h$ for every $h=0,1,2,\dotsc$.
\item
For each $h$ the joint distribution of $r_h \GG$ and $V_h$ is given by
\begin{equation}
\P(r_h\GG=\tau,V_h=v)=\P(r_h\G=\tau) \quad \forall \tau \in \TT{h}, v\in \gen(h,\tau), h\ge 0.
\end{equation}
\item
The joint distribution of $(V_0,V_1,V_2,\dotsc)$ and $\GG$ is determined recursively as follows:
for each $h\ge0$, given $(V_0,V_1,\dotsc,V_h)$ and $r_h \GG$, the numbers of successors $c_v \GG$ are independent as $v$ ranges over $\gen(h,\GG)$, with distribution $p(\cdot)$ for $v\ne V_h$, and with the size-biased distribution $p^*(\cdot)$ for $v=V_h$; given also the numbers of successors $c_v \GG$ for $v \in \gen(h,\GG)$, the vertex $V_{h+1}$ has uniform distribution on the set of $c_{V_h} \GG$ successors of $V_h$.
\end{enumerate}
\end{theorem}

\begin{remark} 
The UICT is a special case for which the critical branching process has off-spring probability $p(n)=(1/2)^{n+1}$. In this case the conditional probability of the left-hand-side of \eqref{yamb:n1} provides the same probability for any tree as well as CT with $n$ vertices and thus defines the uniform measure on this set. The measure on the right-hand side of \eqref{yamb:n1} determines the uniform measure on the set of infinite causal triangulation (UICT).
\end{remark}

\begin{remark}  Another measure of interest is the Gibbs measure on the set of CTs. Its Hamiltonian $H$ is simply the number of triangles multiplied by a coupling $\tilde{\lambda}$ (the ``bare'' cosmological constant). There is a correspondence between the number of triangles and vertices in a tree: let $\tau^{(h)}$ be some finite tree of height $h$ and $t^{(h)}$ its corresponding causal triangulation. The number of triangles in $t^{(h)}$ is equal to $1+ 2\sum_{k=1}^{h-1} Z_k \tau^{(h)} + Z_h\tau^{(h)} = H(t^{(h)})$. The probability of $t^{(h)}$ on the set of casual triangulations of the ``disc'' with height $h$ is given by the Gibbs measure $P_h(t^{(h)}) = Z_h^{-1} e^{-\tilde{\lambda} H(t^{(h)})}$, where $Z_h^{-1}$ is the normalisation. Moreover it is not difficult to prove that for $\tilde{\lambda}=\ln 2$ the measure $P_h$  also converges to the UICT as $h\to \infty$.
\end{remark}

\section{Weak convergence from conditioned critical branching processes}\label{distribution}

Having established the relation between UICTs and critical Galton-Watson processes conditioned to never die out in the previous section, one can now use several known convergence results for the conditioned branching process to determine the corresponding convergence of several observables of the UICT.

From the point of view of universality one expects that the continuum processes shall be the same for any kind of underlying critical Galton-Watson process. We will see that this is indeed the case. Let us therefore consider an arbitrary critical Galton-Watson process $\mathcal{G}$ with generating function $f(s)=\sum_{n\geq0} p(n) s^n$ of the off-spring distribution $p(\cdot)$. Since the process is critical we have $f'(1)=1$. Let us further assume that $\nu=f''(1)/2<\infty$. For short hand denote the size of the $t$'s generation by $\eta_t\equiv Z_t\mathcal{G}$. It was shown by Lindvall \cite{Lindvall1972,Lindvall1974} that if $\eta_0 =\nu t x+ o(t)$ with $x>0$:
\[
\frac{\eta_{[t \tau]}}{\nu t} \codi X_{\tau}, \quad 0\leq \tau <\infty,
\]
where $\codi$ denotes weak convergence on the functions space $D[0,\infty)$ and the continuous process solves the following It\^o's equation
\[
dX_\tau =\sqrt{2 X_\tau} dB_\tau, \quad X_0=x,
\]
with $B_\tau$ standard Brownian motion of variance 1.

Let us note that the finite-dimensional distributions of $\eta_t$ can be easily obtained from the following relation due to Kesten, Ney and Spitzer for the generating function of the size of the $t$'s generation of a critical Galton-Watson process with $\nu=f''(1)/2<\infty$ and $\eta_0=1$ (e.g. see \cite{Athreya1972})
\[
\frac{1}{1-f_t(s)}=\frac{1}{1-s}+\nu t+o(t), \quad \text{uniformly for } 0\leq s<1.
\]
Tightness can then be obtained by standard techniques (e.g. see \cite{Billingsley1999}). An alternative detailed proof of Lindvall's theorem using convergence of the generator of the Markov process can be found in \cite{Either1986}.

We now investigate the convergence of the length of the boundary of an infinite CT as a process of time. Since any Galton-Watson tree conditioned to never die out is in bijection with an infinite CT, we refer to the corresponding probability measure as an infinite CT constructed from a critical Galton-Watson process. The UICT is then a special case for which the critical branching process has off-spring probability $p(n)=(1/2)^{n+1}$. In particular, this off-spring probability satisfies $f^{(n)}(1)<\infty$ for all $n\in\mathbb{N}$.

By the relation discussed in the previous section, the size of the boundary $k_t$ of an infinite CT constructed from a critical Galton-Watson process at time $t$ corresponds to the size of the $t$'s generation of the Galton-Watson process conditioned to never die out, denoted by $\eh_t \equiv Z_t\mathcal{G^\infty}$. Define the length process
\begin{equation} \label{eqlengthprocess}
k^{(t)}_{\tau}:=\frac{k_{[t \tau]}}{\nu t}\equiv\frac{\eh_{[t \tau]}}{\nu t}, \quad 0\leq \tau <\infty,
\end{equation}
The convergence of the finite-dimensional distributions of the process $\{\eh_t \}$ was studied by Lamperti and Ney \cite{Lamperti1968} and we can deduce the following theorem for the length process \eqref{eqlengthprocess}:

\begin{theorem} \label{thmlength}
For an infinite CT constructed from a critical Galton-Watson process with $\nu=f''(1)/2<\infty$ and $f'''(1)<\infty$, and initial boundary $m_0\equiv k_0 =\nu t l+ o(t)$ with $l\geq 0$ we have
\[
k^{(t)}_{\tau} \codi L_\tau, \quad 0\leq \tau <\infty,
\]
in the sense of weak convergence on the functions space $D[0,\infty)$, where the continuous process solves the following It\^o's equation
\[
dL_\tau =2 d\tau +  \sqrt{2 L_\tau} dB_\tau \quad L_0=\ell.
\]
The process $L_\tau$ is diffusive and the Feynman-Kac equation for $\phi_\xi(\ell,\tau)=\E[ \exp(- \xi L_\tau)|L_0=\ell]$ is given by
\[
-\frac{\partial}{\partial \tau} \phi_\xi(\ell,\tau) = \hat{H} \phi_\xi(\ell,\tau), \quad \hat{H}= -2\frac{\partial}{\partial \ell}- \ell\frac{\partial^2}{\partial \ell^2}, \quad \phi_\xi(\ell,0)=e^{-\xi \ell}.
\]
\end{theorem}

Here the operator $\hat{H}$ is known in the physics literature as the Hamiltonian of two-dimensional CDT (having cosmological constant equal zero, see e.g. \cite{Ambjorn:1998xu}).

In \cite{Lamperti1968}, Theorem 1 convergence of the finite-dimensional distributions of the process $\{\eh_t \}$ to those of the above diffusion process was shown. However, to prove convergence of the process one also has to prove tightness. The complete proof of Theorem \ref{thmlength} is presented in Appendix \ref{markov}.

\begin{corollary} \label{length}
In the special case of $l=0$, corresponding to an infinite CT constructed from a critical Galton-Watson process with zero initial boundary, we have
\[
\Ec{ e^{-\xi L_\tau}}{L_0=0} = \frac{1}{(1+\xi \tau)^2}
\]
which for $\tau=1$ is random variable with gamma distribution with parameter two, i.e.\ $\P(\Gamma_n\in dx)/dx=x\,e^{-x}$, $x\geq 0$. (the sum of two independent random variables with exponential distribution with rate $1$).
\end{corollary}

This is gives the distribution of the rescaled upper boundary $L_1$, i.e.\ of the random variable $k_t/t$ in the limit $t\to\infty$. It is hence the analog of Theorem 4 of \cite{Krikun2005} which states the corresponding result for UIPT.

We now want to discuss the convergence of the rescaled area of a neighbourhood of the boundary $\Gamma_t$ of height $t$. Let us denote the number of triangles in $\Gamma_t$ by $\alpha_t$. Define the area process
\begin{equation}\label{areaprocess}
\alpha^{(t)}_{\tau}:=\frac{\alpha_{[t \tau]}}{\nu t^2}, \quad 0\leq \tau <\infty,
\end{equation}
We then have the following theorem based on a theorem of Pakes for conditioned critical Galton-Watson processes \cite{Pakes1999}:

\begin{theorem}\label{thmarea}
For an infinite CT constructed from a critical Galton-Watson process with $\nu=f''(1)/2<\infty$ and $f'''(1)<\infty$, and initial boundary $m_0\equiv k_0 =\nu t l+ o(t)$ we have
\[
(k^{(t)}_{\tau}, \alpha^{(t)}_{\tau}) \codi (L_\tau , 2 \int_0^\tau L_u du ), \quad 0\leq \tau <\infty,
\]
in the sense of weak convergence on the functions space $D[0,\infty)\times D[0,\infty)$, where the continuous process $L_\tau$ solves the It\^o's equation as in Theorem \ref{thmlength}
\[
dL_\tau =2 d\tau +  \sqrt{2 L_\tau} dB_\tau \quad L_0=l.
\]
The Feynman-Kac equation for $\phi_{\xi,\lambda}(l,\tau)=\E[ \exp(- \xi L_\tau - 2 \lambda \int_0^\tau L_u du )|L_0=l]$ is given by
\[
-\frac{\partial}{\partial \tau} \phi_{\xi,\lambda}(l,\tau) = \hat{H} \phi_{\xi,\lambda}(l,\tau), \quad \hat{H}= -2\frac{\partial}{\partial l}- l\frac{\partial^2}{\partial l^2} +2\lambda l, \quad \phi_{\xi,\lambda}(l,0)=e^{-\xi l}.
\]
\end{theorem}

\begin{proof}
By construction of the bijection between CTs and Galton-Watson trees we have $\alpha_t=k_0+2(k_1+...+k_{t-1})+k_t$, i.e. each internal spatial (horizontal) edge is connected to two triangles while each boundary edge is connected to one triangle (see Figure~\ref{f0}). Hence
\begin{equation}\label{atproc}
\alpha^{(t)}_{\tau}=\frac{\alpha_{[t \tau]}}{\nu t^2}=\frac{1}{\nu t^2}\left(2 \sum_{i=0}^{[t \tau]} \eh_i -  \eh_0-\eh_{[t \tau]}\right)=2 \int_0^\tau k_u^{(t)} du + o(1).
\end{equation}
Following ideas of \cite{Pakes1999}, Theorem 3.3, the weak convergence of $(k^{(t)}_{\tau}, \alpha^{(t)}_{\tau})$ then follows from the weak convergence of
\[
(k_{\tau}^{(t)}, 2 \int_0^\tau k_u^{(t)} ) \codi (L_\tau,2 \int_0^\tau L_u du).
\]
It is enough to note that by \eqref{atproc} we have that $h(k^{(t)}_{\tau}):=(k^{(t)}_{\tau}, \alpha^{(t)}_{\tau})$ is a continuous functional of $k^{(t)}_{\tau}$ and hence the convergence of $(k^{(t)}_{\tau}, \alpha^{(t)}_{\tau})\codi (L_\tau,2 \int_0^\tau L_u du)$ follows by the continuous mapping theorem (Theorem 2.7, \cite{Billingsley1999}) applied to Theorem \ref{thmlength}.

Having established the convergence, we can then apply the Feynman-Kac formula to
\[
\phi_{\xi,\lambda}(l,\tau)=\Ec{ e^{- \xi L_\tau - 2 \lambda \int_0^\tau L_u du }}{L_0=l}
\]
with
\[
dL_\tau =2 dt +  \sqrt{2 L_\tau} dB_\tau \quad L_0=l.
\]
which yields
\[
-\frac{\partial}{\partial \tau} \phi_{\xi,\lambda}(l,\tau) =\left(-2\frac{\partial}{\partial l}- l\frac{\partial^2}{\partial l^2} +2\lambda l\right)  \phi_{\xi,\lambda}(l,\tau), \quad \phi_{\xi,\lambda}(l,0)=e^{-\xi l}.
\]
\end{proof}

The last equation is again known from the physics literature in the context of CDT with cosmological constant $\lambda$. In fact, one can easily solve the differential equation leading to
\begin{equation}\label{prop1}
\phi_{\xi,\lambda}(l,\tau) = \frac{\bar{\xi}^2(\xi,\tau)-2\lambda}{\xi^2-2\lambda} e^{-l \bar{\xi}(\xi,\tau)}
\end{equation}
with
\[
\bar{\xi}(\xi,\tau) =\sqrt{2\lambda}\coth(\sqrt{2\lambda}\tau) -\frac{2\lambda}{\sinh^2(\sqrt{2\lambda}\tau) \left[\xi+\sqrt{2\lambda} \coth(\sqrt{2\lambda}\tau)\right]}
\]

\begin{corollary}\label{remarkarea}
Setting $\tau=1$ and $l=0$ in \eqref{prop1} one has
\[
\phi_{\xi,\lambda}(0,1)=\frac{2\lambda}{(\sqrt{2\lambda}\cosh\sqrt{2\lambda}+\xi\sinh\sqrt{2\lambda})^2}
\]
which also follows from \cite{Pakes1999}, Theorem 3.3. In particular, for $\lambda=0$ one recovers
\[
\Ec{ e^{-\xi L_1}}{L_0=0} = \frac{1}{(1+\xi)^2}
\]
as in Corollary \ref{length} and for $\xi=0$
\begin{equation}\label{cosh}
\Ec{ e^{-\lambda A_1}}{L_0=0} = \frac{1}{\cosh^2(\sqrt{2\lambda})}
\end{equation}
with $A_1=2 \int_0^1L_u du$.
\end{corollary}

This gives the distribution of the random variable $A_1$, i.e.\ $\alpha_t/t^2$ in the limit $t\to\infty$.
The distribution of $A_1$ appears at several places related to the study of Brownian motion as has been exposed for example in \cite{Biane}. Based on the discussion in \cite{Biane} we can make two remarks:

\begin{remark}
\label{seriesrep} The random variable $A_1$, as introduced in Remark \ref{remarkarea}, can be written in the following series representation
\[
A_1=\frac{2}{\pi^2}\sum_{n=1}^\infty \frac{\Gamma_{n}}{(n-1/2)^2},
\]
where the $\Gamma_n$, $n\geq 1$ are i.i.d.\ random variables with gamma distribution with parameter two, i.e.\ $\P(\Gamma_n\in dx)/dx=x\,e^{-x}$, $x\geq 0$. The relation can easily be seen by noting that
\begin{equation}\label{seriescosh}
\E(e^{-\lambda \Gamma_n})=\frac{1}{(1+\lambda)^{2}}\quad \text{and} \quad \cosh z=\prod_{n\geq1}\left(1+\frac{z^2}{\pi^{2}(n-1/2)^{2}}\right).
\end{equation}
\end{remark}

\begin{remark}
\label{levy} In the framework of L\'evy-Khintchine representations a distribution is called infinitely divisible iff its Laplace transform $\phi(\lambda)$ admits the following representation
\[
\phi(\lambda)=\exp\left(  -c \lambda -\int_0^\infty (1-e^{-\lambda x})\nu(dx) \right).
\]
for some $c\geq0$. Here $v(dx)$ is the so-called L\'evy measure and for the present application it is sufficient to consider the form of a simple density $v(dx)=\rho(x)dx$.
By a straightforward and explicit computation using \eqref{cosh} and \eqref{seriescosh} one sees that the distribution of $A_1$ is infinitely divisible and has a L\'evy-Khintchine representations with $c=0$ and L\'evy density
\[
\rho(x)=\frac{2}{x}\sum_{n\geq 1}e^{-\pi^2(n-1/2)^2 x/2}.
\]
\end{remark}

\section{Discussion}\label{conclusion}

We discussed infinite causal triangulations and the existence of the uniform measure on those, so-called UICT, in an alternative presentation to \cite{Durhuus:2009sm}. One observes that under this measure the probability of a causal triangulation of a cylinder is related to a critical Galton-Watson process conditioned to never die out. We used this relation to prove weak convergence of the joint rescaled length-area process $(k^{(t)}_\tau,\alpha^{(t)}_\tau)$ of an infinite CT constructed from an arbitrary critical Galton-Watson process to a limiting diffusion process $(L_\tau, A_\tau)$, with $A_\tau=2 \int_0^\tau L_u du$, where the It\^o's equation for $L_\tau$ is given by (e.g.\ Theorem \ref{thmlength} and \ref{thmarea})
\[
dL_\tau =2 d\tau +  \sqrt{2 L_\tau} dB_\tau, \quad L_0=l.
\]
In particular, we show that the Feynman-Kac formula for $\E[\exp(-\xi L_\tau -\lambda A_\tau)| L_0=l]$ corresponds to a imaginary time Schr\"odinger equation with the following Hamiltonian
\[
\hat{H}(l,\partial_l)= -2\frac{\partial}{\partial l}- l\frac{\partial^2}{\partial l^2} +2\lambda l.
\]
This is the well-known Hamiltonian for two-dimensional CDT with cosmological constant $\lambda$ (see \cite{Ambjorn:1998xu}).\footnote{In fact, it is the Hamiltonian acting on an non-rooted boundary. This is due to the fact that by the construction of the Feynman-Kac or Kolmogorov backwards equation we are acting on the upper, non-rooted boundary. Alternatively, one could have also used the Kolmogorov forward equation to obtain the Hamiltonian acting on the rooted, lower boundary.}

By calculating the inverse Laplace transform of \eqref{prop1} one can also obtain the transition amplitude or Green's function
\begin{eqnarray}
\phi_\lambda(l_1,l_2,\tau)&=&\Ec{\mathbb{I}\{L_\tau \in dl_2\}\cdot e^{- 2\lambda  \int_0^\tau L_u du} }{ L_0 = l_1}/dl_2\nonumber\\
&=& \frac{\sL l_2}{\sqrt{l_1 l_2}} \frac{e^{- \sL (l_1+l_2) \coth(\sL \tau)}}{\sinh(\sL\tau)} I_1\left( \frac{2 \sqrt{\lambda l_1 l_2}}{\sinh(\sL\tau)}\right)\nonumber
\end{eqnarray}
where $\mathbb{I}\{\cdot\}$ is the indicator function and $I_1(\cdot)$ is the modified Bessel function of first kind. This expression is also known in physics as the CDT propagator. In particular, setting $\lambda=0$ one obtains the transition amplitude for the length process
\[
\phi_0(l_1,l_2,\tau)= \frac{l_2}{\tau\sqrt{l_1l_2}} e^{-\frac{l_1+l_2}{\tau}} I_1\left( \frac{2\sqrt{l_1l_2}}{\tau}\right)
\]

In conclusion, Theorem \ref{thmlength} and \ref{thmarea} provide us with a mathematical rigorous proof of certain scaling limits of two-dimensional causal dynamical triangulations (CDT). In ongoing work, we further show how to obtain these results in a slightly different manner from a certain growth process of UICT. While in this article we given a mathematical rigorous derivation for several correlations functions of CDT from the UICT it would be interesting to obtain the full scaling limit using a framework like in Le Gall's and Mierment's work on the Brownian map in the context of DT \cite{Clay-Le-Gall}.

We hope that the discussion in the article helps physicists working on quantum gravity, and in particular CDT, to connect their work to the corresponding branching process picture.

\subsection*{Acknowledgments}
The authors would like to thank the anonymous referee, as well as Thordur Jonsson and Sigurdur Stef\'ansson for comments on improvements of the manuscript.
The work of V.S.\ was supported by FAPERJ (grants E-26/170.008/2008 and E-26/110.982/2008) and CNPq (grants 471891/2006-1, 309397/2008-1 and 471946/2008-7). The work of A.Y.\ was partly supported by CNPq 308510/2010-0. S.Z.\ would like to thank the Department of Statistics at S\~ao Paulo University (IME-USP) as well as the Institute for Pure and Applied Mathematics (IMPA) for kind hospitality. Financial support of FAPESP under project 2010/05891-2, as well as STFC and EPSRC is kindly acknowledged.
\appendix

\section{Proof of Theorem \ref{thmlength}}\label{markov}

Define $\nu=f''(1)/2<\infty$ as before and $\mu=f'''(1)/2<\infty$. Recall that we want to show convergence of
\begin{equation}
k^{(t)}_{\tau}=\frac{\eh_{[t \tau]}}{\nu t} \codi L_\tau, \quad 0\leq \tau <\infty,
\end{equation}
on the functions space $D[0,\infty)$. To do so we consider the rescaled process
\begin{equation}\label{Bconv}
\tilde{k}^{(t)}_{\tau}= \nu k^{(t)}_{\tau} \codi \tilde{L}_\tau= \nu L_{\tau}
\end{equation}
where then $\tilde{L}_\tau$ is a diffusion process with generator
\begin{equation}
Ag(x)=2\nu g'(x)+\nu x g''(x),
\end{equation}
where by Theorem 2.1 of Chapter 8 of \cite{Either1986} one has $g\in C_c^\infty([0,\infty))$, i.e.\ the set of continouse functions $f: [0,\infty)\to\mathbb{R}$ which are infinitely differentiable and have compact support in $[0,\infty)$. To show convergence of the process $\tilde{k}^{(t)}_{\tau}$ to the diffusion $\tilde{L}_\tau$ with the above generator we follow closely the strategy employed in Theorem 1.3 of Chapter 9 in \cite{Either1986} to prove Lindvall's theorem.

Note that $\eh_{n}/t$ is a Markov chain taking values in $E_t=\{l/t | l=1,2,3,...\}$. Given $\eh_{n}=tx$ we can then write
\begin{equation} \label{appBeta}
\eh_{n+1}= \sum_{k=1}^{tx-1}\xi_k + \xi_0,
\end{equation}
where $\xi_k$ for $k\geq 0$ are iid random variables with generating function $f(s)$ and $\xi_0$ is a random variable with generating functions $sf'(s)$. Recall that $f(s)$ is the generating function of the off-spring probabilities. The above statement follows directly from Theorem \ref{t:big}. Indeed, by Theorem \ref{t:big} we have
\begin{equation}
\sum_{k\geq0} \P(\eh_{n+1}=k| \eh_{n}=tx ) s^k =\frac{1}{tx} s\frac{d}{ds} f^{tx}(s)= f^{tx-1}(s) \cdot sf'(s)
\end{equation}
which is the generating function for \eqref{appBeta}.
We have
\begin{eqnarray}
\E \xi_k &=&  1, \quad \E \xi_k^2 = 1+2 \nu, \quad \text{for $k\geq 1$} \nonumber\\
\E \xi_0 &= &1+2 \nu, \quad \E \xi_0^2 = 1+6 \nu +2 \mu
\end{eqnarray}
We now define
\begin{equation}
T_t g(x)=\E \left\{g\left( \frac{1}{t}\left[ \sum_{k=1}^{tx-1}\xi_k + \xi_0\right]  \right)\right\}.
\end{equation}
By Theorem 6.5 of Chapter 1 and Corollary 8.9 of Chapter 4 of \cite{Either1986}, to prove the convergence \eqref{Bconv} it is enough to show that
\begin{equation}
\lim_{t\to\infty} \sup_{x\in E_t} | t(T_t g(x) -g(x)) - 2\nu g'(x) -\nu x g''(x)|=0,\quad g\in C_c^\infty([0,\infty)).
\end{equation}

For $x\in E_t$ we define
\begin{eqnarray}
\varepsilon_t (x)&=& t(T_t g(x) -g(x)) - 2\nu g'(x) -\nu x g''(x)\nonumber\\
 &=& \E\left\{ t g\left(\frac{1}{t}\left[ \sum_{k=1}^{tx-1}\xi_k +\xi_0 \right] \right) -t g(x) -(\xi_0-1) g'(x) +\right.\nonumber\\
 &&\left. -\frac{1}{2}g''(x) \frac{1}{t}\left[ \left(\sum_{k=1}^{tx-1}(\xi_k-1)\right)^2 + \xi_0 -1 \right]   \right\}\nonumber\\
 &=& \Delta^{1}_t(x)+\Delta^{2}_t(x)
\end{eqnarray}
where
\begin{eqnarray}
\Delta^{1}_t(x) &=& \frac{1} {t} \mu g''(x) \\
\Delta^{2}_t(x) &=& \E\left\{ \int_0^1S^2_{tx} x(1-u)\left[ g''(x+u \sqrt{\frac{x}{t}}S_{tx} )-g''(x) \right] du\right\}
\end{eqnarray}
and
\begin{equation}
S_{tx}=\frac{1}{\sqrt{tx}} \left(  \sum_{k=1}^{tx-1} (\xi_k-1) + (\xi_0-1) \right).
\end{equation}
Now, since $\mu=f'''(1)/2<\infty$ and 
we have
\begin{equation}
\lim_{t\to\infty} \sup_{x\in E_t} |\Delta^{1}_t(x)| \leq \mu \|g''\| \lim_{t\to\infty} \frac{1}{t}=0.
\end{equation}
Let us suppose that $g$ has support in $[0,c]$. Then, since $\xi_k\geq0$ for $k\geq1$ and $\xi_0\geq1$ we have
\begin{equation}
x+u \sqrt{\frac{x}{t}}S_{tx} \geq x(1-u).
\end{equation}
hence one gets that
\begin{eqnarray}
&&\left| \int_0^1S^2_{tx} x(1-u)\left[ g''(x+u \sqrt{\frac{x}{t}}S_{tx} )-g''(x) \right] du\right| \nonumber\\
&\leq& 2x S_{tx}^2 \int_{0 \vee (1-c/x)}^1 (1-u) \|g'' \|du =  x \|g'' \| ((c/x) \wedge 1)^2 S_{tx}^2 \label{Binequ}
\end{eqnarray}

To show that $\lim_{t\to\infty} \sup_{x\in E_t} |\Delta^{2}_t(x)|=0$ it suffices to show that  one has  $\lim_{t\to\infty} |\Delta^{2}_t(x_t)|=0$ for any convergent series $x_t$, as well as for $x_t\to0$ and $x_t\to\infty$. Let us first treat the special cases $\lim_{t\to\infty} x_t =0$ and $\lim_{t\to\infty} x_t=\infty$. Note that
\begin{equation}
\E S^2_{tx}= 2\nu +\frac{2\mu}{t x}\leq 2(\nu+\mu), \quad \text{for all $t$ and } x\in E_t. \label{BS2bound}
\end{equation}
From \eqref{Binequ} and \eqref{BS2bound} it then follows that $\lim_{t\to\infty}| \Delta^{2}_t(x_t)|=0$ if  $\lim_{t\to\infty} x_t =0$ or $\lim_{t\to\infty} x_t=\infty$.

We now consider the case $\lim_{t\to\infty} x_t =x$, where $0<x<\infty$. In this case one has
\begin{equation}
\lim_{t\to\infty} \E e^{r S_{tx_t}}=e^{\nu r^2}
\end{equation}
and hence $S_{tx_t}\codi \Sigma$ with $\Sigma\sim\mathcal{N}(0,2\nu)$. Following the steps of Theorem 1.3 in \cite{Either1986} Chapter 9, one then obtains $\lim_{t\to\infty}| \Delta^{2}_t(x_t)|=0$ from \eqref{Binequ} and the dominant convergence theorem.

Hence we showed that
\begin{equation}
\lim_{t\to\infty} \sup_{x\in E_t} |\varepsilon_t (x)| =0.
\end{equation}
Noting that the initial condition converges $\tilde{k}_0^{t} \rightarrow \nu l$ with $l\geq0$ one completes the proof. $\Box$

\providecommand{\href}[2]{#2}\begingroup\raggedright\endgroup

\end{document}